\documentclass[seceq]{ptptex}

\usepackage{graphicx}
\usepackage{wrapft}

\title{
Growing length scales during aging in 2$d$ disordered systems
}

\author{
H. Rieger, G. Schehr and R. Paul
}

\inst{
Theoretische Physik, Universit\"at des Saarlandes, 66041 Saarbr\"ucken, Germany
}

\abst{
The non-equilibrium dynamics of three paradigmatic models for
two-dimensional systems with quenched disorder is studied with a
focus on the existence and analysis of a growing length scale during aging
at low temperatures: 1) The random bond Ising ferromagnet, 2) the
Edwards-Anderson model for a spin glas, 3) the solid-on-solid model on
a disordered substrate (equivalent to the sine-Gordon model with
random phase shifts). Interestingly, we find in all three models a
length scale that grows algebraically with time (up to the system size
in cases 1 and 3, up to the finite equilibrium length in case 2) with a
temperature dependent growth exponent. Whereas in cases 1 and 2 this
length scale characterizes a coarsening process, it represents in
case 3 the growing size of fluctuations during aging. 
}

\begin{document}

\maketitle

\section{Introduction}

The non-equilibrium dynamics of disordered, in particular of glassy
systems has become a very rich field in recent years and
despite many efforts the understanding of non-equilibrium dynamics of
disordered and glassy systems in finite dimensions remains a
challenging problem. In particular in glasses and spin glasses the
aging process displays a very rich phenomenology demanding new
theoretical concepts \cite{leshouches}. But already less complex
--- and apparently less glassy --- systems, like disordered but
non-frustrated systems \cite{random-ising} or even pure systems 
\cite{cugliandolo_pure} reveal interesting and
unexpected aging phenomena. One of the most intriguing questions in
this context is whether the out-of-equilibrium dynamics is essentially
fully determined by a coarsening process (a question that even arises
in the more complex spin glass situation \cite{nao-rie-rev}),
describable by a growing length scale that characterizes essentially
all out-of-equilibrium processes. In this paper we will consider three
paradigmatic models for two-dimensional systems with quenched disorder
with a focus on existence and analysis of a growing length
scale during aging at low temperatures: the random bond Ising
ferromagnet, the Edwards-Anderson (EA) model for a spin glass, and the
solid-on-solid model on a disordered substrate which is equivalent to
the sine-Gordon model with random phase shifts.

\section{The random bond Ising Ferromagnet}

As the first example for two-dimensional disordered system we consider
the random bond Ising ferromagnet. It is defined by the Hamiltonian
\begin{equation}
\label{rbim}
H = - \sum\limits_{(ij)} J_{ij} S_i S_j, \quad S_i = \pm 1,
\end{equation}
where the couplings $J_{ij}$ are non-negative quenched random
variables of variance $\varepsilon$ and the sum is over all nearest
neighbor pairs $(ij)$ on a square lattice of size $L\times L$ with
periodic boundary conditions.  This paradigmatic model for a
disordered magnetic system (with bond- or temperature randomness) with
an Ising symmetry has a second order phase transition from a
paramagnetic to a ferromagnetic phase at a critical temperature
$T_c(\epsilon)$ that decreases with increasing disorder strength
$\epsilon$. For temperatures $T$ below $T_c$ the magnetization
$\overline{\langle m_i\rangle_T}$, where $\langle\cdots\rangle_T$
means the thermal average and $\overline{\cdots}$ the average over the
disorder, takes on a non-vanishing value. 

Non-equilibrium dynamics at temperatures below $T_c$ arises for
instance via an instantaneous quench of the systems from the
paramagnetic phase to a temperature below $T_c$. A stochastic process
defined by single spin-flip transition rates defined for instance by
the Metropolis rules
$w(S_i\to-S_i)=1/(1+\exp(-\beta(H(S_i)-H(-S_i))))$ models a
non-conserved order parameter dynamics and can be studied by computer
simulations. For a quench below $T_c$ the dynamics is a coarsening
process during which ferromagnetic domains of a typical lateral
extension $R(t)$ form, where $t$ is the time elapsed after the quench
(see Fig. 1). A standard way to extract this time dependent length
scale is via the spatial two-point correlation function
$C(r,t)=\overline{\langle m_i(t)m_{i+r}(t)\rangle_T}$, which is 
expected to scale like $C(r,t)=\tilde{c}(r/R(t))$.

\begin{figure}[t]
\centerline{\includegraphics*[width=\columnwidth]{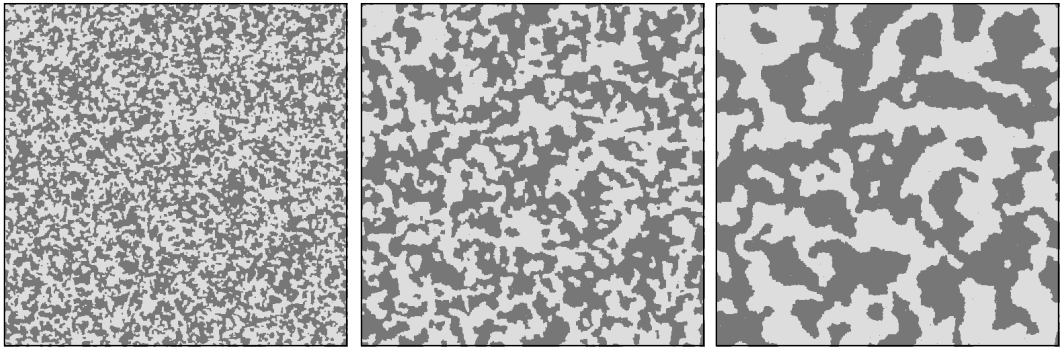}}
\caption{Domain growth in the RBIM with Glauber kinetics. We show evolution
pictures at $t=10^2$, $10^4$ and $10^6$ MCS for a $512 \times 512$
lattice, after a quench from $T = \infty$ to $T = 0.5$ and $J_{ij}$
uniformly distributed between 0 and 2 ($J_{ij}\in[0,2]$). The up spins
are marked in black, and the down spins are marked in grey.}
\label{fig1}
\label{fm-evol}
\end{figure}

\begin{figure}[t]
\includegraphics*[width=0.5\columnwidth]{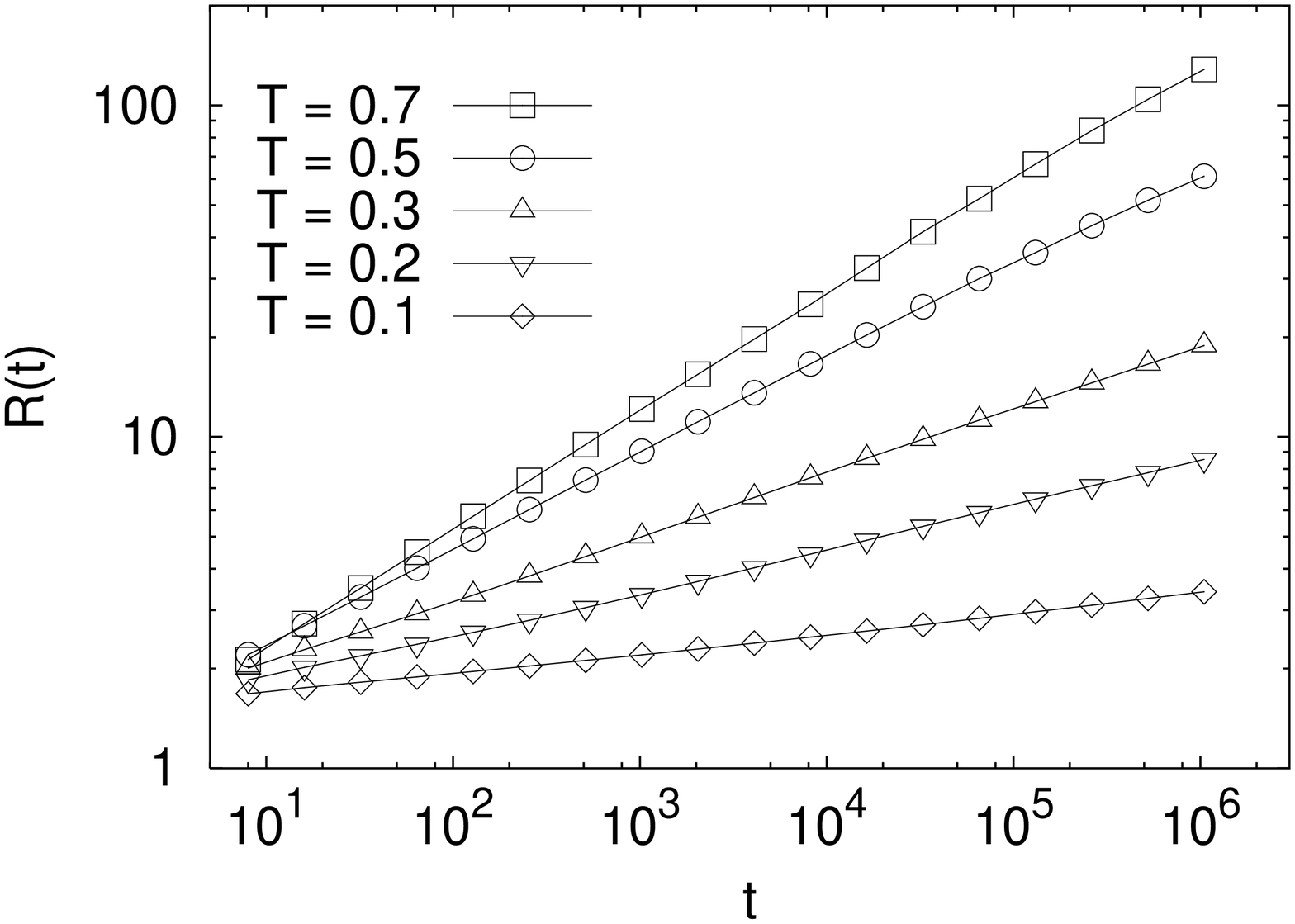}
\includegraphics*[width=0.5\columnwidth]{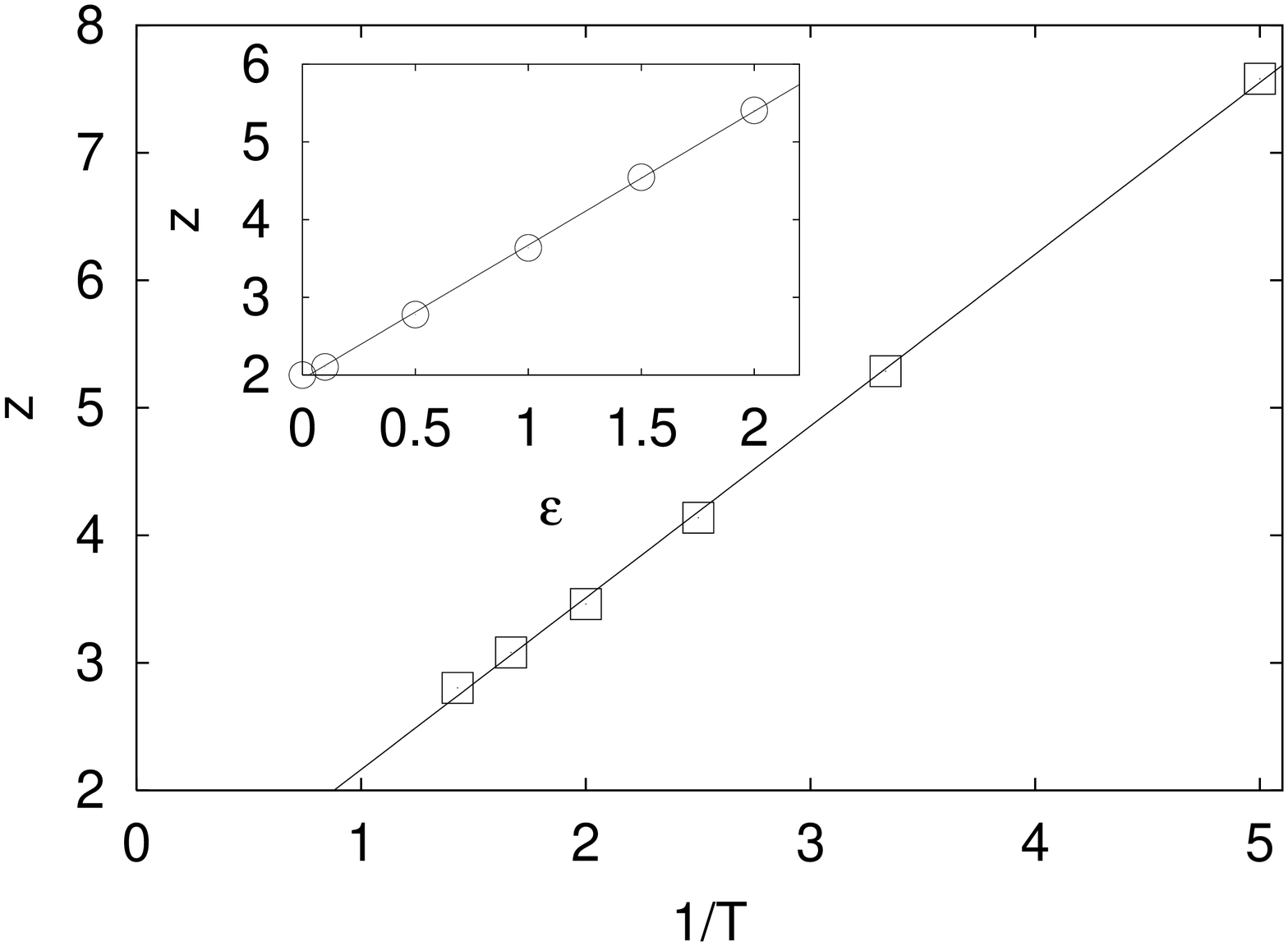}
\caption{{\bf Left:} 
Log-log plot of the correlation length $R$ vs. time $t$ in the 2$d$
random bond Ising model for different temperatures and
$J_{ij}\in[0,2]$. {\bf Right:} Estimates of the exponent $z$ vs. $1/T$
for the data shown left (note that for $T>1\approx T_c$, i.e.\ for
$1/T<1$ the system is in the paramagnetic phase). Inset shows $z$
vs. the disorder strength $\epsilon$ for fixed temperature $T=0.5$
(the distribution $P(J)$ is chosen to be uniform over
$[1-\epsilon/2,1+\epsilon/2])$.  The straight lines (in the main figure
and in the inset) represent the analytical prediction $z=2+\epsilon/T$.}
\label{rbfm-length}
\end{figure}

An important study of the non-conserved random bond Ising model (RBIM) is due to Huse and
Henley (HH) \cite{hh85}. HH argued that coarsening
domains are trapped by energy barriers $E_B(R) \simeq E_0 R^{\psi}$,
with exponent $\psi = \chi/(2- \zeta)$, where
$\chi$ and $\zeta$ are the pinning and roughening exponents. For $d=2$, these
exponents are known to be $\chi = 1/3$ and $\zeta=2/3$ \cite{fns77},
yielding $\psi=1/4$. As a consequence of the HH scenario one expects
the following scaling scenario for the length scale $R(t)$:
\begin{equation}
R(t)/R_0=h(t/t_0)\;,{\rm with}\quad h(x)\sim
\left\{\begin{array}{lcl}
x^{1/2}   & {\rm for} & x\ll1\\
(\ln x)^4 & {\rm for} & x\gg1
\end{array}\right.
\label{HH}
\end{equation}
where $R_0\sim T^4$ and $t_0\sim T^8$. Instead we find (via an
extensive Monte-Carlo study, see \cite{rbim}) that $R(t)$ grows
algebraically with a temperature and disorder strength dependent
exponent $1/z(T,\epsilon)$:
\begin{equation}
R(t)\sim t^{1/z(T,\epsilon)}\;{\rm for}\quad t\gg t_0
\quad{\rm with}\quad z(T,\epsilon)=2+\epsilon/T
\label{power}
\end{equation}
where the time $t_0$ does not depend on $T$ and $\epsilon$ (see
Fig. \ref{rbfm-length}).  This algebraic growth law with a temperature
and disorder strength dependent growth exponent $\theta$ indicates a
logarithmic barrier scaling form $E_B(R)\sim\epsilon\ln(1+R)$ in
contrast to the algebraic form $E_B\propto R^\psi$ assumed in the HH
picture.

\section{2$d$ EA spin glass}

Here we consider the two-dimensional Ising spin glass with
nearest-neighbor interactions distributed according to a Gaussian
with zero mean and variance one
\begin{equation}
  H=-\sum_{\langle ij \rangle} J_{ij} S_i S_j \;,
  {\rm with}\quad
  P(J_{ij})=\frac{1}{\sqrt{2\pi}}\exp\left( -\frac{J_{ij}^2}{2}\right).  
\label{hami}
\end{equation}
This model is in a paramagnetic state for all temperature $T>0$ but
displays a very slow dynamics at low temperatures which can be
observed for instance in the non-equilibrium dynamics occurring after a
quenched from high temperatures. It turns out \cite{sg-2d} that this
aging process can be characterized by a coarsening process up to a
maximum domain size given by the equilibrium correlation length
$\xi_{\rm eq}$.

\begin{figure}[t]
\centerline{\includegraphics*[width=\columnwidth]{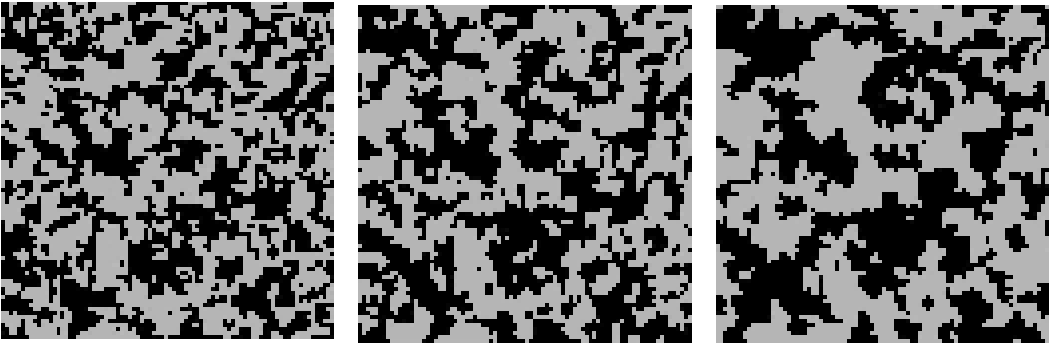}}
\caption{Domain growth in the 2$d$ EA model with Gaussian couplings
for $T=0.3$. The system size is $L=100$, {\it i.e.} much smaller than in
Fig. \ref{fm-evol}. The snapshots show the domains relative to the
ground state after $t=10^2$, $10^4$ and $10^6$ Monte-Carlo sweeps. }
\label{sg-evol}
\end{figure}

It is possible to calculate exactly the ground state (GS) of this system
using for instance a minimal weight perfect matching algorithm
\cite{rieger-opt}.  Denoting the GS for a particular
disorder realization with $\{S_i^0\}$ we define the local overlap with
it as $q_i^{gs}(t)=S_i(t)S_i^0$. For a ferromagnetic system (i.e.\
$J_{ij}=J>0$) the GS obviously has $S_i^0=1$ and therefore $q_i$
corresponds to the (time dependent) local magnetization.
In Fig.\ \ref{sg-evol} snapshots of the time evolution of the local GS
overlap are depicted, showing an increasing average domain size. In
contrast to the time evolution of a random bond ferromagnet shown in
Fig. \ref{fm-evol} even for very large waiting times very small domains
exist. These are either very stable clusters because strong bonds have
to be broken to flip the spins or new domains within the bigger ones
appear since less strongly bound spins initialize the formation of a
new domain.

\begin{figure}[b]
\centerline{
\includegraphics*[width=0.45\columnwidth]{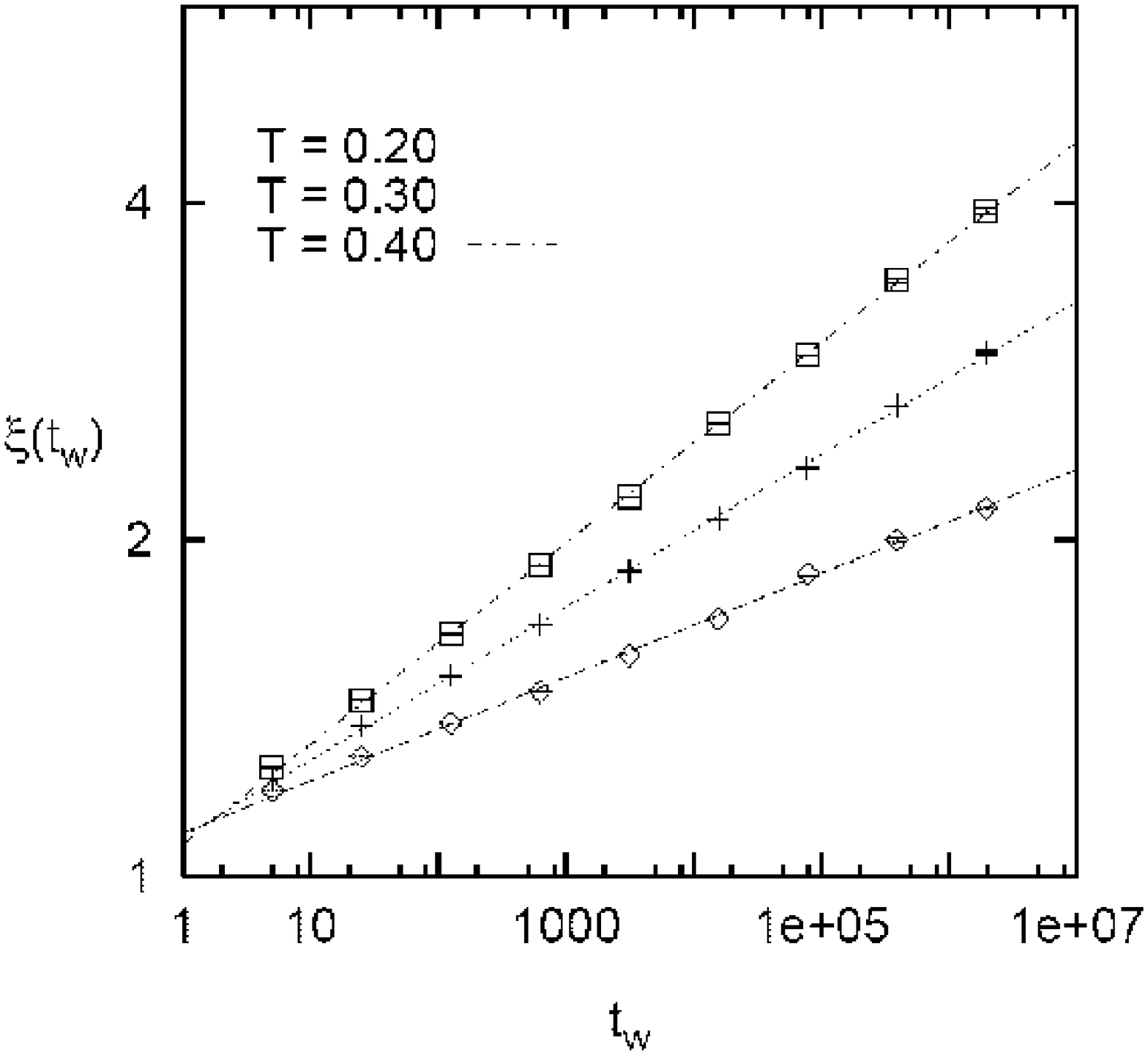}
\includegraphics*[width=0.45\columnwidth]{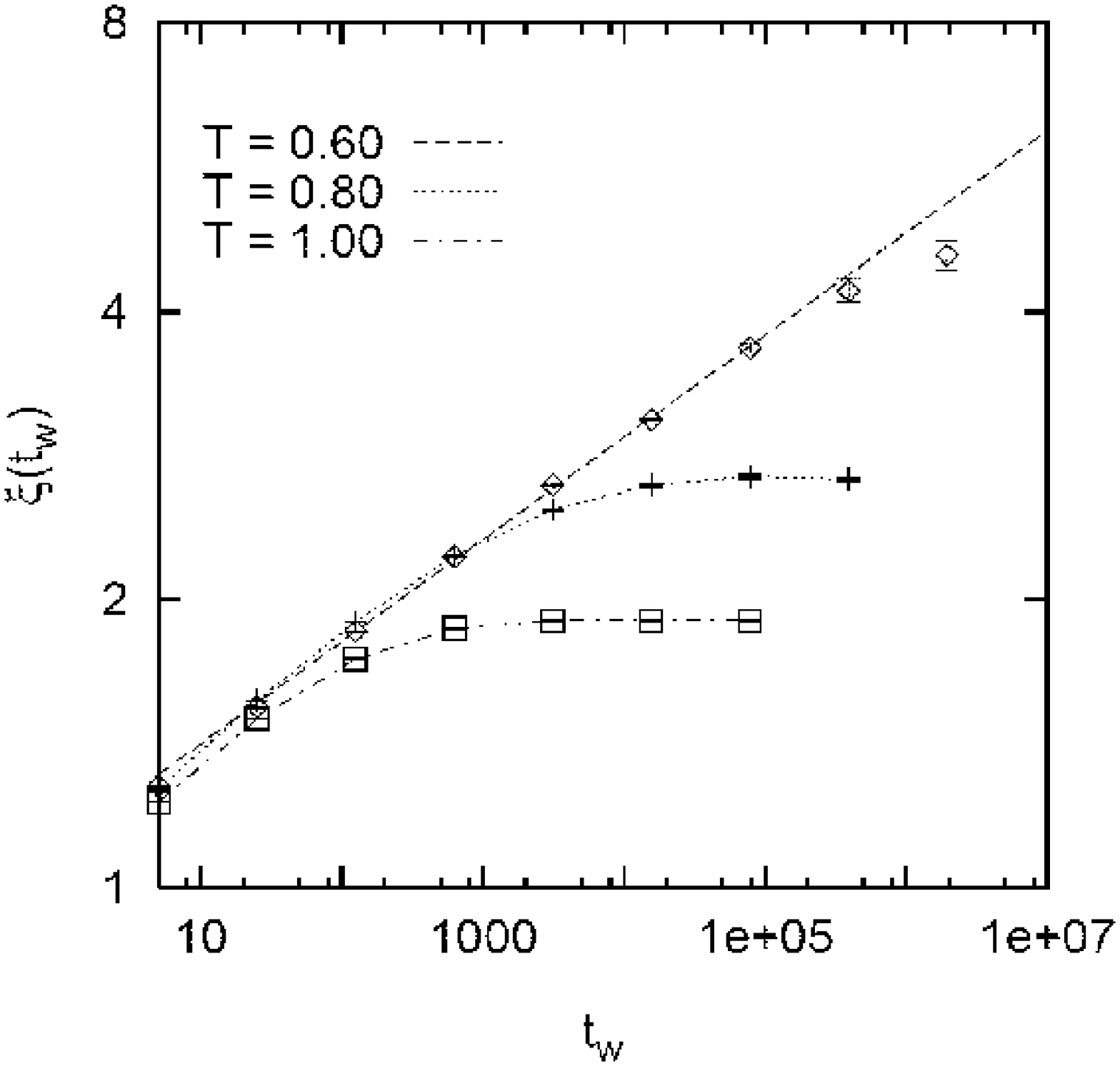}
}
\caption{Correlation length in the 2$d$ EA SG model as a function 
of time $t_w$ for different temperatures. {\bf Left:} Fits to an
algebraic growth law $\xi(t_w)\sim t^{1/z(T)}$ with
$1/z(T=0.2)=0.046$, $1/z(T=0.3)=0.068$ and
$1/z(T=0.4)=0.090$.  {\bf Right:} For sufficiently large
temperatures the time dependent correlation length $\xi(t)$
saturates at the equilibrium correlation length $\xi_{\rm eq}(T)$
within the accessible time window.}
\label{sg-xi}
\end{figure}

The spatial correlation function 
\begin{equation}
G(r,t)=\overline{\langle q_i^{gs}(t) q_{i+r}^{gs}(t)\rangle}
\end{equation}
allows for a quantitative analysis of the domain size evolution. It
turns out that it scales like $G(r,t)=g(r/\xi(t))$ and we can obtain
an estimate for the correlation length (or typical domain size) via an
integral of $G(r,t)$ over $r$. The result is shown in
Fig. \ref{sg-xi}. Note that for increasing temperatures a) the domain
growth speeds up, b) the equilibrium correlation length gets
smaller. As a consequence of both tendencies one can observe the
saturation of the time dependent correlation length at the finite
equilibrium correlation length for higher temperatures on the right panel.

We observe that the data for $\xi(t)$ (in the non-equilibrium regime
$\xi(t)\ll\xi_{\rm eq}(T)$) can very well be fitted by an algebraic
growth law with a temperature dependent exponent $z(T)$:
\begin{equation}
\xi(t)\propto t^{1/z(T)}\quad{\rm with}\quad z(T)\approx (0.23\cdot T)^{-1}
\end{equation}
which displays again the $1/T$ behavior that we have encountered
already in the last section for the random bond ferromagnet,
indicating also here the presence of logarithmic barriers.

\section{2$d$ SOS model on a disordered substrate}

Here, we investigate the non
equilibrium relaxational dynamics of a solid-on-solid (SOS) model on a
disordered substrate, defined on a two dimensional square lattice and
described by the following elastic Hamiltonian in terms of height
variables $h_i$
\begin{eqnarray}
H_{\text{SOS}} = \sum_{\langle ij \rangle} (h_i - h_j)^2 \quad , \quad
 h_i \equiv n_i + d_i,
 \label{Def_SOS} 
\end{eqnarray}
where $n_i$ are unbounded discrete variables, {\it i.e.} $n_i \in \{0,
\pm 1, \pm 2,..\}$ and $d_i \in [0,1[$ are uniformly distributed quenched
random offsets, uncorrelated from site to site. In the absence of
disorder, {\it i.e.} $d_i=0$, the model exhibits a roughening
transition in the same universality class as the Kosterlitz-Thouless
transition \cite{sos-pure}, at a temperature $T_r$ separating a flat phase
at low $T$ from a logarithmically (thermally) rough one above
$T_r$. The presence of disorder is known to modify significantly the
nature of the transition \cite{sos-rand}. The so-called {\it
superroughening} transition occurs at a temperature $T_g = T_r/2 =
2/\pi$. Above $T_g$, where the disorder is irrelevant on large length
scales, the surface is logarithmically rough again, although below
$T_g$ the system exhibits a glassy phase where the pinning disorder
induces a stronger roughness of the interface.

\begin{figure}[t]
\includegraphics[width=0.5\columnwidth]{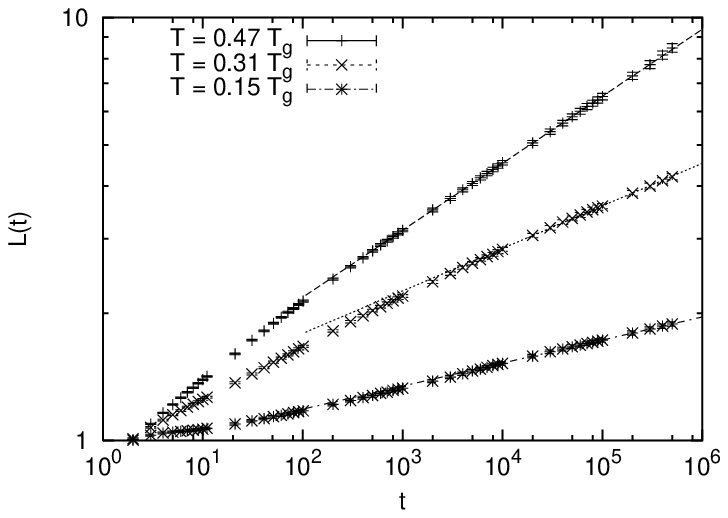}
\includegraphics[width=0.5\columnwidth]{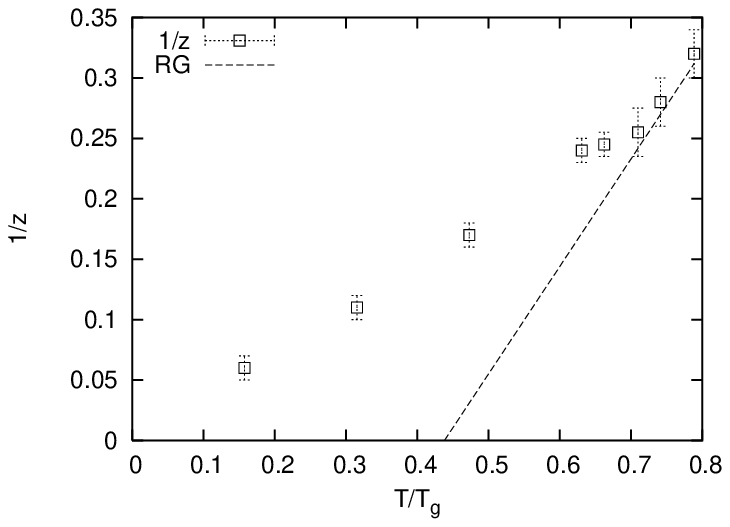}
\caption{{\bf Left:} Growing length scale ${\cal L}(t)$
 for different temperatures. The solid lines are guides to the
 eyes. {\bf Right:} $1/z(T)$ as a function of $T/T_g$. The dashed line
 which shows the result of the one loop RG is drawn without any
 fitting parameter.  }
\label{sos-length}
\end{figure}

The spatial (2-point) connected correlation function is defined as
\begin{eqnarray}
C(r,t) = \frac{1}{L^2} \sum_i \overline{\langle h_i(t) h_{i+r}(t) \rangle - 
\langle h_i(t) \rangle \langle h_{i+r}(t) \rangle}
\label{Def_Struct_Fact} 
\end{eqnarray}
which scales as
\begin{eqnarray}
C(r,t) = {\cal F}\left(r/{\cal L}(t)\right) \quad{\rm with} \quad
 {\cal L}(t) \sim t^{1/z}\;. \label{Scaling_Struct_Fact}
\end{eqnarray}
Therefore one can estimate ${\cal L}(t)$ by integrating $C(r,t)$ over
$r$.  In Fig. \ref{sos-length} we show the value of ${\cal L}(t)$
computed in this way for different temperatures. One obtains a rather
good fit by a power law ${\cal L}(t)\sim t^{1/z(T)}$, thus obtaining a
value of the $T$ dependent dynamical exponent. One notices also that
${\cal L}(t)$ approaches an algebraic growth after a pre-asymptotic
regime which increases with decreasing temperature.
Figure \ref{sos-length} shows our estimate for $1/z(T)$ as a function of
$T$ (for details see \cite{rieger-schehr}. As expected, the dynamical
exponent is a decreasing function of the temperature. One expects that
$z=2$ for $T>T_g$ and that it becomes $T$-dependent below $T_g$ with
$z = 2 + 2e^{\gamma_E}\tau + {\cal O}(\tau^2)$ as predicted by a one
loop RG calculation
\cite{sos-loop}.  At high temperature $T>T_g$ and in the
vicinity of $T_g^-$, it is numerically rather difficult to extract a
reliable estimate for the dynamical exponent due to finite size
effects. Therefore we restrict ourselves here to lower
temperatures $T < 0.8 \,T_g$. For temperature $T\gtrsim 0.7 \,T_g$, the
value of $z$ is still in reasonable agreement with the RG
prediction. Around the value $T^*
\simeq 0.63 \,T_g$, where $z \simeq 4$, the curve $1/z(T)$
shows an inflection point, below which $1/z$ decreases linearly with
$T$. In this regime, $z(T)$ is well fitted by
\begin{eqnarray}
z(T) \approx 4\cdot T^*/T 
\quad{\rm for}~T\le T^*\approx0.63T_g.
\label{z_1oT}
\end{eqnarray}     
This behavior $z \propto 1/T$ is compatible with an activated dynamics
over logarithmic barriers, {\it i.e.} an Arrhenius type behavior
$t_{\text{typ}} \sim e^{{B_L}_{\text{typ}}/T}$ with
$B_{L_{\text{typ}}} \sim \log{L_{\text{typ}}}$.

One would like to relate the length scale ${\cal L}(t)$ to the size of
spatially correlated structures like domains or droplets.  We first
explored the idea that at low temperature, the nonequilibrium dynamics
could be understood as a coarsening process reflected in a spatially
growing correlation with the ground state (GS).  Interestingly,
computing the GS of the SOS model on a disordered substrate
(\ref{Def_SOS}) is a minimum cost flow problem for which exits a
polynomial algorithm and can therefore be computed exactly
\cite{sos-gs}.  After determining one GS
$n_i^0$ (note that the GS, which is computed with free boundary
conditions, is infinitely degenerated since a global shift of all
heights by an arbitrary integer is again a GS), we define for each
time $t$ the height difference $m_i(t) = n_i(t) - n_i(0)$ and identify
the connected clusters (domains) of sites with identical $m_i(t)$
using a depth-first search algorithm. Notice that for the comparison
with the ground state, the Monte Carlo simulations are performed here
using free boundary conditions.

\begin{figure}[t]
\centering
\includegraphics[width=\columnwidth]{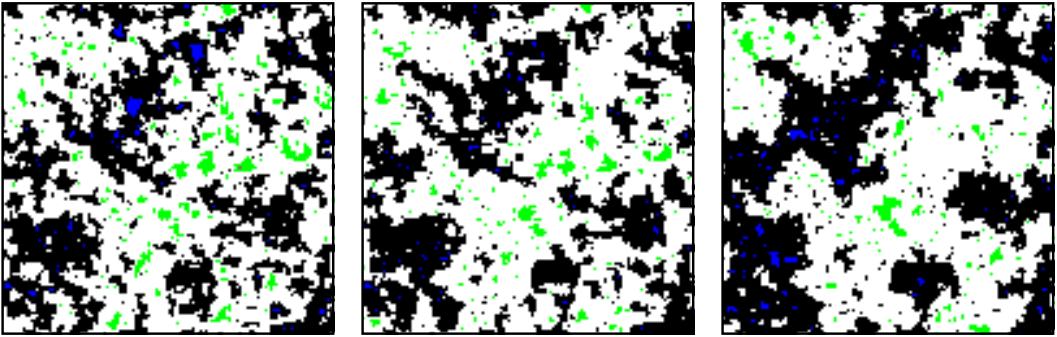}    
\caption{
 Snapshot of the height field of the random SOS model relative to the
 ground state $m_i(t) = n_i(t) - n_i^0$ for $T=0.47\,T_g$. The system
 site is $L=128$. Different colors correspond to different values of
 $m_i(t)$ : $m_i(t) = -2$ in green, $m_i(t) = -1$ in white, $m_i(t) =
 0$ in black and $m_i(t)=+1$ in blue and so on.  Note that large
 domains in white and black persist and change only slowly in time.
 }\label{CoarsPict}
\end{figure}

In Fig. \ref{CoarsPict} we show snapshots of these domains for
$T<T_g$. Starting from a random initial configuration one can for
$T<T_g$ very quickly ($t\lesssim100$) identify large domains that
evolve only very slowly at later times. On the other hand for $T>T_g$
the configurations decorrelated very quickly in time. To make this
analysis more quantitative, we determined the cluster size
distribution $P_{\text{th}}(S,t)$ for one realization of the disorder
(and for different realizations of the thermal noise).

\begin{figure}[b]
\centering
\includegraphics[width=0.48\columnwidth]{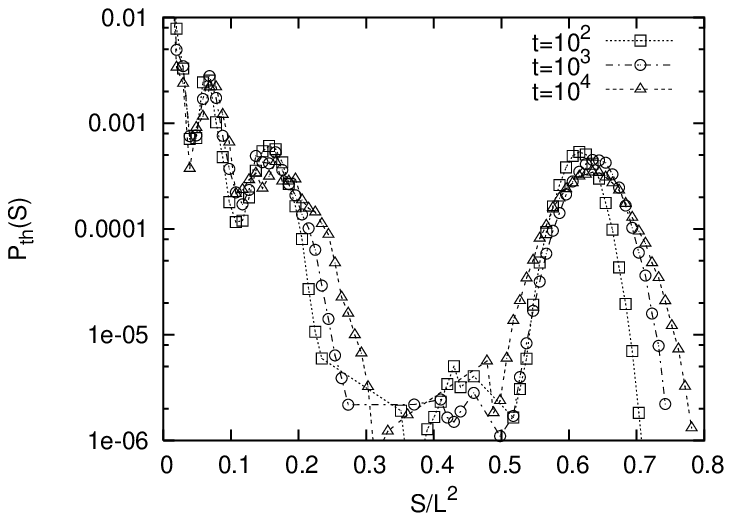}   
\includegraphics[width=0.48\columnwidth]{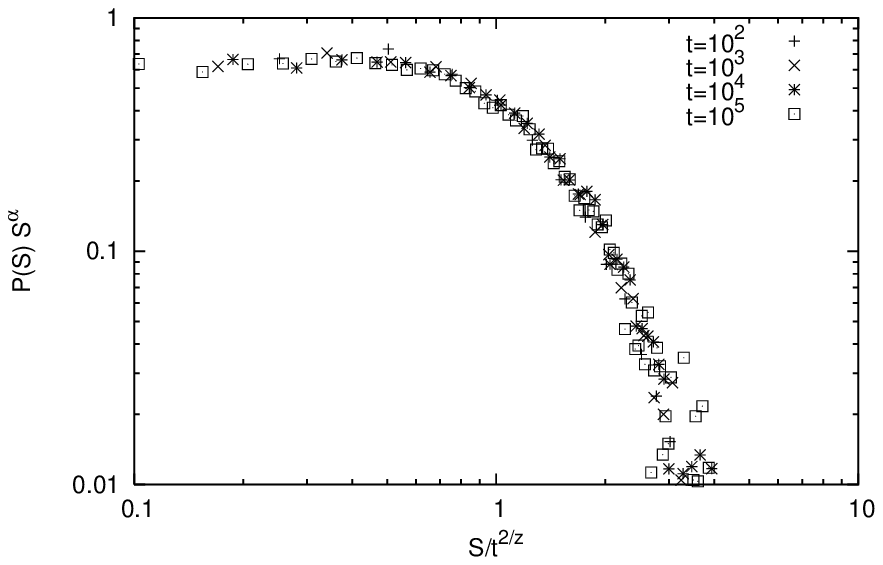}    
\caption{{\bf Left:} Size distribution $P_{\text{th}}(S,t)$ (see definition in the
 text) for different times $t$. Here $T = 0.47 T_g$. {\bf Right:}
 $S^{\alpha} P^{\text{GS}}_{\text{droplet}}(S,t)$ with $\alpha = 1.9
 \pm 0.1$ as a function of $S/t^{2/z}$ with $2/z = 0.26 \pm
 0.03$. Here the initial condition is the ground state and $T=0.3
 \,T_g$.}
\label{RescDroplet_0.2} 
\label{Dist_Therm}
\end{figure}

As shown on Fig. \ref{Dist_Therm}, $P_{\text{th}}(S,t)$ starts to
develop a peak at a rather large value $S^*(t)$ on the earlier stage of
the dynamics (this peak also develops if we start with a random
initial configuration). It turns out that $S^*(t)$ is the size of the
largest connected flat cluster of the ground state configuration
$n_i^0 = C^{\text{st}}$. On the time scales presented here, as time
$t$ is growing, this peak remains stable $S^*(t) \simeq
C^{\text{st}}$, implying that the system is {\it not} coarsening.
At later times, as suggested by simulations on
smaller systems, this peak progressively disappears and the
distribution becomes very flat. We also checked that the mean size of
these connected clusters is not directly related to ${\cal L}(t)$.
One has however to keep in mind that we are computing the {\it
connected} correlation functions, 
{\it i.e.} we measure the thermal fluctuations of the height profile
around its mean (typical) value $\langle h_i(t) \rangle$. Therefore, we
believe that these connected correlations are instead related to the
broadening of this ``stable'' peak (Fig. \ref{Dist_Therm}), {\it i.e.} the
fluctuations around this typical state at time $t$.

To characterize more precisely the fluctuations around this
cluster, we identify ``droplets'' by initializing the
system in the ground state itself $n_i(t=0) = n_i^0$. At low temperature, 
and on the time scales explored here, one expects that 
the ground state represents a good
approximation of a typical configuration, {\it i.e.} $\langle n_i(t)
\rangle \simeq n_i^0$. We compute the distribution
$P^{\text{GS}}_{\text{droplet}}(S,t)$ of the sizes of the connected
clusters with a common value of $m_i(t)\ne0$. It turns out, as shown
in Fig. \ref{RescDroplet_0.2}, that
$P_{\text{droplet}}^{\text{GS}}(S,t)$ obeys the scaling form
\begin{equation}
P_{\text{droplet}}^{\text{GS}}(S,t) = \frac{1}{S^{\alpha}} 
{\cal F}_{\text{droplet}}^{\text{GS}} 
\left( \frac{S}{{\cal L}^2(t)}\right) \quad, \quad \alpha = 1.9 \pm 0.1, 
\label{active_GS}  
\end{equation}
where $\alpha$ is independent of $T$ within the accuracy of our data
and ${\cal L}(t) \sim t^{1/z}$. The value of $z$ in (\ref{active_GS})
is in good agreement with the one extracted from the $2$-point
correlation function $C(r,t) = F(r/{\cal L}(t))$
(\ref{Scaling_Struct_Fact}).

\section{Conclusion}

We studied the time dependent correlation length $R(t)$ in three
models of disordered systems in two dimensions that are characterized
by distinct features: 1) The random bond Ising ferromagnet as an
example for a random system that has long range order at low
temperatures and is expected to perform a simple (but slow) coarsening
process after a temperature quench into the ordered phase. 2) The
Edwards-Anderson spin glass model as an example for a frustrated
system without a finite temperature critical point and an ordered
phase but with an extremely slow dynamics and a large correlation
length at very low temperatures. 3) The disordered SOS model as a
model with a critical point and a low temperature phase without long
range order but infinite correlation length. Surprisingly, in spite of
the pronounced differences between these systems we find that all
three show an algebraic dependence of the correlation length on the
age $t$ of the system $R(t)\sim t^{1/z(T)}$ and that the exponent $z$
(which would be identical to the dynamical exponent if the system is
critical) depends linearly on the inverse temperature:
\begin{equation}
z(T)\propto 1/T\;.
\label{zt}
\end{equation}
If the dynamics at low temperatures in all three systems is thermally
activated, this behavior hints at a logarithmic scaling behavior of the
energy barriers as a function of their size. In the disordered SOS
model one would actually expect such a scaling \cite{rieger-schehr}.
For the EA spin glass (in 2$d$) the situation is complicated by the fact
that the ground state is not expected to be stable with respect to
thermal fluctuations, i.e. in principle excitations of
increasing size would cost less and less energy --- therefore
a logarithmic barrier scaling comes a bit as a surprise.
Finally for the random bond Ising model a simple scaling picture
\cite{hh85} based on the scaling behavior of the domain walls 
in this model would predict an algebraic energy scaling --- resulting
in a formally infinite value for $z$, which is not confirmed by our
results \cite{rbim}. Hence we have to conclude with the observation
that the common behavior (\ref{zt}) of the growth exponent $z$ in 2$d$
disordered models indicates a more complicated and yet hidden
mechanism that is active in the non-equilibrium dynamics of these
systems at low temperatures at least during the first 10 decades of
the aging process.

We also would like to emphasize the fact that the physical
interpretation of the growing length scale in the three systems under
consideration in this paper is quite different: In the random bond
ferromagnet it is simply the typical transverse domain size, where
domains are easily identified as connected clusters of common
magnetization sign. In the EA spin glass model the length scale is
also determined by a domain size -- where domains are defined as
connected clusters of spins with common orientation with respect to
one of the two ground states. These domains grow steadily up to a
maximum size set by the equilibrium correlation length. In the
disordered SOS model, however, the growing length scale is {\it not}
connected to growing domains --- actually the system settles quite
fast after the temperature quench into a configuration that has a
pretty large overlap with one of the ground states. Instead of growing
further these initially very large domains thermal fluctuations of
increasing size destroy these domains --- and it is the spatial extent
of these fluctuations that is characterized by the growing length
scale studied here.

\section*{Acknowledgements}
  G. Schehr acknowledges the financial support provided through the
  European Community´s Human Potential Program under contract
  HPRN-CT-2002-00307, DYG\-LAGEMEM, and R. Paul's work was 
  supported by the DFG (SFB277).

%

\end{document}